\documentclass[preprint2]{aastex}

\usepackage{verbatim, graphicx, natbib, ifthen, pbox}
\usepackage{eso-pic}
\newboolean{ColorPlots}
\newcommand{\asec}{\prime\prime}
\newcommand{\QR}{2001~QR$_{322}$}
\newcommand{\QO}{2014~QO$_{441}$}
\newcommand{\QP}{2014~QP$_{441}$}

\shorttitle{Observation of Two New L4 Neptune Trojans}
\shortauthors{Gerdes et al.}

\begin{document}

\title{Observation of Two New L4 Neptune Trojans in the Dark Energy Survey Supernova Fields}

\author{
D.~W.~Gerdes\altaffilmark{1},
R.~J.~Jennings\altaffilmark{2},
G.~M.~Bernstein\altaffilmark{3},
M.~Sako\altaffilmark{3},
F.~Adams\altaffilmark{1,4},
D.~Goldstein\altaffilmark{5,6},
R.~Kessler\altaffilmark{7,8},
T.~Abbott\altaffilmark{9},
F.~B.~Abdalla\altaffilmark{10},
S.~Allam\altaffilmark{11},
A.~Benoit-L{\'e}vy\altaffilmark{10},
E.~Bertin\altaffilmark{12,13},
D.~Brooks\altaffilmark{10},
E.~Buckley-Geer\altaffilmark{11},
D.~L.~Burke\altaffilmark{14,15},
D.~Capozzi\altaffilmark{16},
A.~Carnero~Rosell\altaffilmark{17,18},
M.~Carrasco~Kind\altaffilmark{19,20},
J.~Carretero\altaffilmark{21,22},
C.~E.~Cunha\altaffilmark{14},
C.~B.~D'Andrea\altaffilmark{16},
L.~N.~da Costa\altaffilmark{17,18},
D.~L.~DePoy\altaffilmark{23},
S.~Desai\altaffilmark{24,25},
J.~P.~Dietrich\altaffilmark{24,26},
P.~Doel\altaffilmark{10},
T.~F.~Eifler\altaffilmark{3,27},
A.~Fausti Neto\altaffilmark{17},
B.~Flaugher\altaffilmark{11},
J.~Frieman\altaffilmark{11,7},
E.~Gaztanaga\altaffilmark{21},
D.~Gruen\altaffilmark{28,26},
R.~A.~Gruendl\altaffilmark{19,20},
G.~Gutierrez\altaffilmark{11},
K.~Honscheid\altaffilmark{29,30},
D.~J.~James\altaffilmark{9},
K.~Kuehn\altaffilmark{31},
N.~Kuropatkin\altaffilmark{11},
O.~Lahav\altaffilmark{10},
T.~S.~Li\altaffilmark{23},
M.~A.~G.~Maia\altaffilmark{17,18},
M.~March\altaffilmark{3},
P.~Martini\altaffilmark{29,32},
C.~J.~Miller\altaffilmark{4,1},
R.~Miquel\altaffilmark{22,33},
R.~C.~Nichol\altaffilmark{16},
B.~Nord\altaffilmark{11},
R.~Ogando\altaffilmark{17,18},
A.~A.~Plazas\altaffilmark{27},
A.~K.~Romer\altaffilmark{34},
A.~Roodman\altaffilmark{14,15},
E.~Sanchez\altaffilmark{35},
B.~Santiago\altaffilmark{36,17},
M.~Schubnell\altaffilmark{1},
I.~Sevilla-Noarbe\altaffilmark{35,19},
R.~C.~Smith\altaffilmark{9},
M.~Soares-Santos\altaffilmark{11},
F.~Sobreira\altaffilmark{11,17},
E.~Suchyta\altaffilmark{29,30},
M.~E.~C.~Swanson\altaffilmark{20},
G.~Tarl\'e\altaffilmark{1},
J.~Thaler\altaffilmark{37},
A.~R.~Walker\altaffilmark{9},
W.~Wester\altaffilmark{11},
Y.~Zhang\altaffilmark{1}
\\ \vspace{0.2cm} (The DES Collaboration) \\
}
 
\altaffiltext{1}{Department of Physics, University of Michigan, Ann Arbor, MI 48109, USA}
\altaffiltext{2}{Carleton College, Northfield, MN 55057, USA}
\altaffiltext{3}{Department of Physics and Astronomy, University of Pennsylvania, Philadelphia, PA 19104, USA}
\altaffiltext{4}{Department of Astronomy, University of Michigan, Ann Arbor, MI 48109, USA}
\altaffiltext{5}{Department of Astronomy, University of California, Berkeley,  501 Campbell Hall, Berkeley, CA 94720, USA}
\altaffiltext{6}{Lawrence Berkeley National Laboratory, 1 Cyclotron Road, Berkeley, CA 94720, USA}
\altaffiltext{7}{Kavli Institute for Cosmological Physics, University of Chicago, Chicago, IL 60637, USA}
\altaffiltext{8}{Department of Astronomy and Astrophysics, University of Chicago, 5640 South Ellis Avenue, Chicago, IL 60637, USA}
\altaffiltext{9}{Cerro Tololo Inter-American Observatory, National Optical Astronomy Observatory, Casilla 603, La Serena, Chile}
\altaffiltext{10}{Department of Physics \& Astronomy, University College London, Gower Street, London, WC1E 6BT, UK}
\altaffiltext{11}{Fermi National Accelerator Laboratory, P. O. Box 500, Batavia, IL 60510, USA}
\altaffiltext{12}{CNRS, UMR 7095, Institut d'Astrophysique de Paris, F-75014, Paris, France}
\altaffiltext{13}{Sorbonne Universit\'es, UPMC Univ Paris 06, UMR 7095, Institut d'Astrophysique de Paris, F-75014, Paris, France}
\altaffiltext{14}{Kavli Institute for Particle Astrophysics \& Cosmology, P. O. Box 2450, Stanford University, Stanford, CA 94305, USA}
\altaffiltext{15}{SLAC National Accelerator Laboratory, Menlo Park, CA 94025, USA}
\altaffiltext{16}{Institute of Cosmology \& Gravitation, University of Portsmouth, Portsmouth, PO1 3FX, UK}
\altaffiltext{17}{Laborat\'orio Interinstitucional de e-Astronomia - LIneA, Rua Gal. Jos\'e Cristino 77, Rio de Janeiro, RJ - 20921-400, Brazil}
\altaffiltext{18}{Observat\'orio Nacional, Rua Gal. Jos\'e Cristino 77, Rio de Janeiro, RJ - 20921-400, Brazil}
\altaffiltext{19}{Department of Astronomy, University of Illinois, 1002 W. Green Street, Urbana, IL 61801, USA}
\altaffiltext{20}{National Center for Supercomputing Applications, 1205 West Clark St., Urbana, IL 61801, USA}
\altaffiltext{21}{Institut de Ci\`encies de l'Espai, IEEC-CSIC, Campus UAB, Carrer de Can Magrans, s/n,  08193 Bellaterra, Barcelona, Spain}
\altaffiltext{22}{Institut de F\'{\i}sica d'Altes Energies, Universitat Aut\`onoma de Barcelona, E-08193 Bellaterra, Barcelona, Spain}
\altaffiltext{23}{George P. and Cynthia Woods Mitchell Institute for Fundamental Physics and Astronomy, and Department of Physics and Astronomy, Texas A\&M University, College Station, TX 77843,  USA}
\altaffiltext{24}{Excellence Cluster Universe, Boltzmannstr.\ 2, 85748 Garching, Germany}
\altaffiltext{25}{Faculty of Physics, Ludwig-Maximilians University, Scheinerstr. 1, 81679 Munich, Germany}
\altaffiltext{26}{Universit\"ats-Sternwarte, Fakult\"at f\"ur Physik, Ludwig-Maximilians Universit\"at M\"unchen, Scheinerstr. 1, 81679 M\"unchen, Germany}
\altaffiltext{27}{Jet Propulsion Laboratory, California Institute of Technology, 4800 Oak Grove Dr., Pasadena, CA 91109, USA}
\altaffiltext{28}{Max Planck Institute for Extraterrestrial Physics, Giessenbachstrasse, 85748 Garching, Germany}
\altaffiltext{29}{Center for Cosmology and Astro-Particle Physics, The Ohio State University, Columbus, OH 43210, USA}
\altaffiltext{30}{Department of Physics, The Ohio State University, Columbus, OH 43210, USA}
\altaffiltext{31}{Australian Astronomical Observatory, North Ryde, NSW 2113, Australia}
\altaffiltext{32}{Department of Astronomy, The Ohio State University, Columbus, OH 43210, USA}
\altaffiltext{33}{Instituci\'o Catalana de Recerca i Estudis Avan\c{c}ats, E-08010 Barcelona, Spain}
\altaffiltext{34}{Department of Physics and Astronomy, Pevensey Building, University of Sussex, Brighton, BN1 9QH, UK}
\altaffiltext{35}{Centro de Investigaciones Energ\'eticas, Medioambientales y Tecnol\'ogicas (CIEMAT), Madrid, Spain}
\altaffiltext{36}{Instituto de F\'\i sica, UFRGS, Caixa Postal 15051, Porto Alegre, RS - 91501-970, Brazil}
\altaffiltext{37}{Department of Physics, University of Illinois, 1110 W. Green St., Urbana, IL 61801, USA}

\AddToShipoutPictureBG*{%
  \AtPageUpperLeft{%
    \hspace{0.75\paperwidth}%
    \raisebox{-3.5\baselineskip}{%
      \makebox[0pt][l]{\textnormal{DES 2015-0052}}   
}}}%

\AddToShipoutPictureBG*{%
  \AtPageUpperLeft{%
    \hspace{0.75\paperwidth}%
    \raisebox{-4.5\baselineskip}{%
      \makebox[0pt][l]{\textnormal{Fermilab PUB-15-299-AE}}   
}}}%

\begin{abstract}
 We report the discovery of the eighth and ninth known Trojans in stable orbits around Neptune's leading Lagrange point, L4. 
The objects \object{ \QO}\ and \object{\QP}\ were detected in data obtained during the 2013-14 and 2014-15 observing seasons by the Dark Energy Survey, using the Dark Energy Camera (DECam) on the 4-meter Blanco telescope at Cerro Tololo Inter-American 
Observatory. Both are in high-inclination orbits (18.8$^{\circ}$ and 19.4$^{\circ}$ respectively). With an eccentricity of 0.104, \QO\ has the most eccentric orbit of the eleven known stable 
Neptune Trojans. Here we describe the search procedure and investigate
the objects' long-term dynamical stability and physical properties. 
\end{abstract}

\maketitle

\section{Introduction}

Trojan asteroids, together with main belt asteroids and members of the classical Kuiper Belt, constitute the only dynamically stable populations of minor planets in the solar system. Trojans share the same orbital period as a major planet, leading or trailing the major planet by approximately 60 degrees in an orbit centered upon the L4 or L5 Lagrange point. Trojan asteroids are associated with a number of planets, including Earth \citep{Connors2011}. Jupiter has the largest and best-characterized population, with over 6000 known Trojans. The total number of  $>$1~km-sized Jupiter Trojans is estimated to exceed 600,000 \citep{Yoshida2005}, comparable to the similar-sized population of the main asteroid belt \citep{Jewitt2000}. The number of large ($\gtrsim 65$~km) Neptune Trojans may exceed their Jovian counterparts by more than an order of magnitude \citep{Chiang2005, Sheppard2006}. 

Since the discovery of the first Neptune Trojan, \QR\ \citep{Chiang2003}, only eight additional Neptune co-orbitals have been discovered prior to this work.\footnote{The Minor Planet Center lists a 9th object, 2004~KV$_{18}$, as an L5 Trojan, but with an eccentricity of 0.18 and a libration amplitude of $\sim 70^{\circ}$, it is known to be unstable on $\sim$Myr timescales and is likely a temporarily captured scattered disk object \citep{Nesvorny2002, Horner2012, Guan2012}.} Seven of the nine previously known Neptune Trojans occupy the L4 region. There is no \textit{a priori} reason to expect the L4 and L5 populations
to be particularly different \citep{Sheppard2010a}; the dearth of known L5 Trojans can be ascribed to the fact that Neptune's L5 region is presently on a line of sight to the galactic 
center, a crowded field of point-like sources against which the detection of transients is observationally challenging \citep{Sheppard2010a, Parker2013}.

In addition to their intrinsic interest, the population of Neptune Trojans provides an important set of constraints on the dynamical
evolution of our Solar System. Although these bodies have relatively little total mass, they play the role of a canary in the coal mine: such small bodies act as test particles and are easily disrupted
through a variety of channels, especially in the early epochs of Solar System history.  Possible mechanisms that could affect the orbits of
these bodies include gas drag in the early solar nebula \citep{Peale1993, Murray1994}, planetary migration \citep{Gomes1998, Kortencamp2004}, the growing
mass of planets during their formation \citep{Marzani1998, Fleming2000}, encounters with passing stars in the solar
birth cluster (which can easily perturb Neptune's orbit, see \citealt{Li2015}), and a wide range of resonance phenomena \citep{Morbidelli2005}.
 Determining the properties and orbital characteristics of the Trojan population also places interesting constraints on theories of
their formation \citep{Chiang2005}. For example, \citet{Parker2015} uses the high mean inclination of Neptune Trojans to argue that the disk into which Neptune migrated must have
been  dynamically excited prior to Neptune's arrival. 

In this paper, we report the discovery of the eighth and ninth known L4 Neptune Trojans, \object{\QO}\ and \object{\QP}. This paper is organized as follows. Section~\ref{sec:overview} briefly describes the Dark Energy Survey and camera, DECam. In Section~\ref{sec:observations} we describe the supernova data reduction pipeline that is the source of transient candidates considered in this analysis, discuss our method for identifying distant solar system objects, and present our observations of the two Neptune Trojans. In Section~\ref{sec:stability} we describe 
numerical investigations of their long-term dynamical stability. Section~\ref{sec:physical} describes their physical characteristics in the context of the broader population of Trojans. We conclude in Section~\ref{sec:summary} with a summary of our results and a discussion of their implications. 

\section{The Dark Energy Survey}
\label{sec:overview}

The Dark Energy Survey (DES; \citealt{Flaugher2005}) is a five-year optical imaging survey being carried out on the 4-meter Blanco telescope
at Cerro Tololo Inter-American Observatory in Chile. Observations are carried out with the Dark Energy Camera (DECam; \citealt{Flaugher2015}), a 3 deg$^{2}$ 
prime-focus camera whose focal plane consists of 62 2k$\times$4k fully-depleted, red-sensitive CCDs. DECam saw first light in September 2012, and 
underwent  ``Science Verification'' operations between November 2012 and February 2013. DECam's power as a discovery instrument for distant 
solar system objects was demonstrated with the discovery, during community time early in the commissioning phase, of the Sedna-like dwarf planet 2012~VP$_{113}$ \citep{Trujillo2014}, which has the most distant perihelion ($>80$~AU) of any known solar system object. Full-scale survey operations began in August 2013 and will
continue through at least 2018.

To achieve the goal of measuring the dark energy equation of 
state to high precision, DES is divided into two distinct, interleaved surveys. The Wide Survey covers 5000 deg$^{2}$ of the south galactic cap in the $grizY$ bands, imaging each survey tile approximately ten times in each band over the course of the survey in order to perform high-statistics measurements of weak gravitational lensing, galaxy-galaxy correlations, and properties of galaxy clusters. The DES Supernova Program (DES-SN, \citet{Bernstein2012}) images ten distinct DECam fields (8 ``shallow'' and 2 ``deep'') in the $griz$ bands at approximately weekly intervals throughout the DES observing season, which runs from mid-August through mid-February.  Characteristics of these fields are shown in Table~\ref{tab:SNfields}. Although primarily intended to find and characterize large numbers of Type Ia supernovae, the supernova fields are also excellent hunting grounds for distant minor planets, which move slowly enough that they can remain in the same field of view for weeks or months at a time, and even from one DES observing season to the next. In particular, the two Stripe-82 fields (``S-fields'') and the three XMM-LSS fields (``X fields'') have moderate ecliptic latitudes ranging from $-20$ to $-15$ degrees, and also, serendipitously, lead the present position of Neptune by approximately 60 degrees of ecliptic longitude. We search for Neptune Trojans in these fields.  

\begin{deluxetable}{llcccc}
\tabletypesize{\scriptsize}
\tablecaption{Summary of DES supernova fields. \label{tab:SNfields}}
\tablewidth{0pt}
\tablehead{
\colhead{} & \colhead{} & \multicolumn{2}{c}{Center (equatorial)} & \multicolumn{2}{c}{Center (ecliptic)} \\
\colhead{Field} & \colhead{Depth}  & \colhead{RA (hours)} & \colhead{DEC (deg.)} & \colhead{$\lambda$ (deg.)} & \colhead{$\beta$ (deg.)} 
}
\startdata
C1 & shallow & 03:37:05.83 & -27:06:41.8 & 42:49:30.2 & -44:52:29.6 \\ 
C2 & shallow & 03:37:05.83 & -29:05:18.2 & 41:52:47.4 & -46:44:19.2 \\ 
C3 & deep & 03:30:35.62 & -28:06:00.0 & 40:25:56.8 & -45:19:21.4 \\ 
X1 & shallow & 02:17:54.17 & -04:55:46.2 & 30:28:17.9 & -17:38:53.3 \\ 
X2 & shallow & 02:22:39.48 & -06:24:43.6 & 31:06:23.4 & -19:26:48.6 \\ 
X3 & deep & 02:25:48.00 & -04:36:00.0 & 32:31:58.0 & -18:00:28.3 \\ 
S1 & shallow & 02:51:16.80 & 00:00:00.0 & 40:22:15.4 & -15:41:10.2 \\ 
S2 & shallow & 02:44:46.66 & -00:59:18.2 & 38:26:48.0 & -16:07:37.7 \\ 
E1 & shallow & 00:31:29.86 & -43:00:34.6 & 346:05:17.0 & -41:44:05.3 \\ 
E2 & shallow & 00:38:00.00 & -43:59:52.8 & 346:43:33.9 & -43:11:58.4 \\ 
\enddata
\end{deluxetable}

\section{Observations and Search Strategy}
\label{sec:observations}

This analysis uses imaging data processed for the Supernova Survey by the DES data management (DESDM) pipeline \citep{Mohr2012, Desai2012} at the
National Center for Supercomputing Applications (NCSA). After cleanup and detrending, astrometric solutions are obtained using \textsc{scamp} \citep{Bertin2006}, with stellar positions in each image matched to the UCAC4 catalog \citep{Zacharias2012} to an absolute astrometric precision of roughly 
0.1$^{\asec}$. Image subtraction is performed with the \textsc{Hotpants} algorithm \citep{hotpants} using PSF-matched deep templates obtained during Science Verification. Source detection is performed on the 
subtracted images using \textsc{Sextractor} \citep{Bertin1996}. A machine-learning algorithm \citep{Goldstein2015} reduces artifacts from instrumental or image-subtraction effects by  a factor of 
nine. All data have been reprocessed using the difference-imaging pipeline described in \citet{Kessler2015}. The efficiency of the pipeline for reconstructing and accepting true point-like sources is measured by inserting fake supernovae into the images using the \textsc{snana} code \citep{Kessler2009} prior to image-subtraction.  The efficiency reaches 50\% for signal-to-noise ratio (SNR)
of about 5 (see Fig 7 in \citet{Kessler2015}), which corresponds to an average magnitude depth of 23.5 in the shallow fields
and 24.5 in the deep fields. We compute PSF magnitudes in the $AB$ system. Images are calibrated nightly to tertiary standards to a photometric precision of 2\% on average.

The $r$- and $i-$bands have the best S/N, with $5\sigma$ limiting magnitudes of 
23.8 in the shallow fields and 24.5 in the deep fields. We therefore carry out the initial candidate search in these bands, and add observations from $g$ and $z$-band once
candidates have been identified. Our target fields contain approximately 185,000 single-epoch transients in the $ri$ bands from the first (Aug. 2013 to Feb. 2014) and second (Aug. 2014 to Feb. 2015) DES observing seasons. 

An object at 30~AU undergoing retrograde motion moves at an apparent rate of up to 4$^{\asec}$ per hour. In the shallow fields, exposure times range from 150~s in $r$-band to 400~s in $z$-band. These exposures are sufficiently short that such an object appears stellar and is accepted with high efficiency by
the machine-learning cut. In the deep fields, however, image subtraction and source detection are ordinarily performed on co-added sequences that range from $3\times 200$~s in $g$-band to $11\times 360$~s in $z$-band. Because such a procedure would frequently result in the rejection of slow-moving transients, for the deep fields we have run a modified version of the transient detection pipeline that uses the single-epoch exposures. This procedure has the added advantage of producing a fine-grained time series well-suited for light-curve measurements as discussed in Section~\ref{sec:physical}.

The apparent motion of a distant solar system object in the roughly one week between DES visits to each field is due largely to Earth's reflex motion. 
Beginning with the list of transients identified in a given visit by the difference-imaging pipeline, we search for counterparts at two subsequent visits in a window consistent with seasonally-appropriate reflex motion at a rate of less than 150$^{\asec}$/day. Triplets selected via this geometrical technique are tested for goodness of fit to an orbit using 
code derived from the work of \citet{Bernstein2000}. Observations from other bands and visits are added iteratively. 

\begin{figure}[htbp]
\begin{center}
\includegraphics[width=\columnwidth]{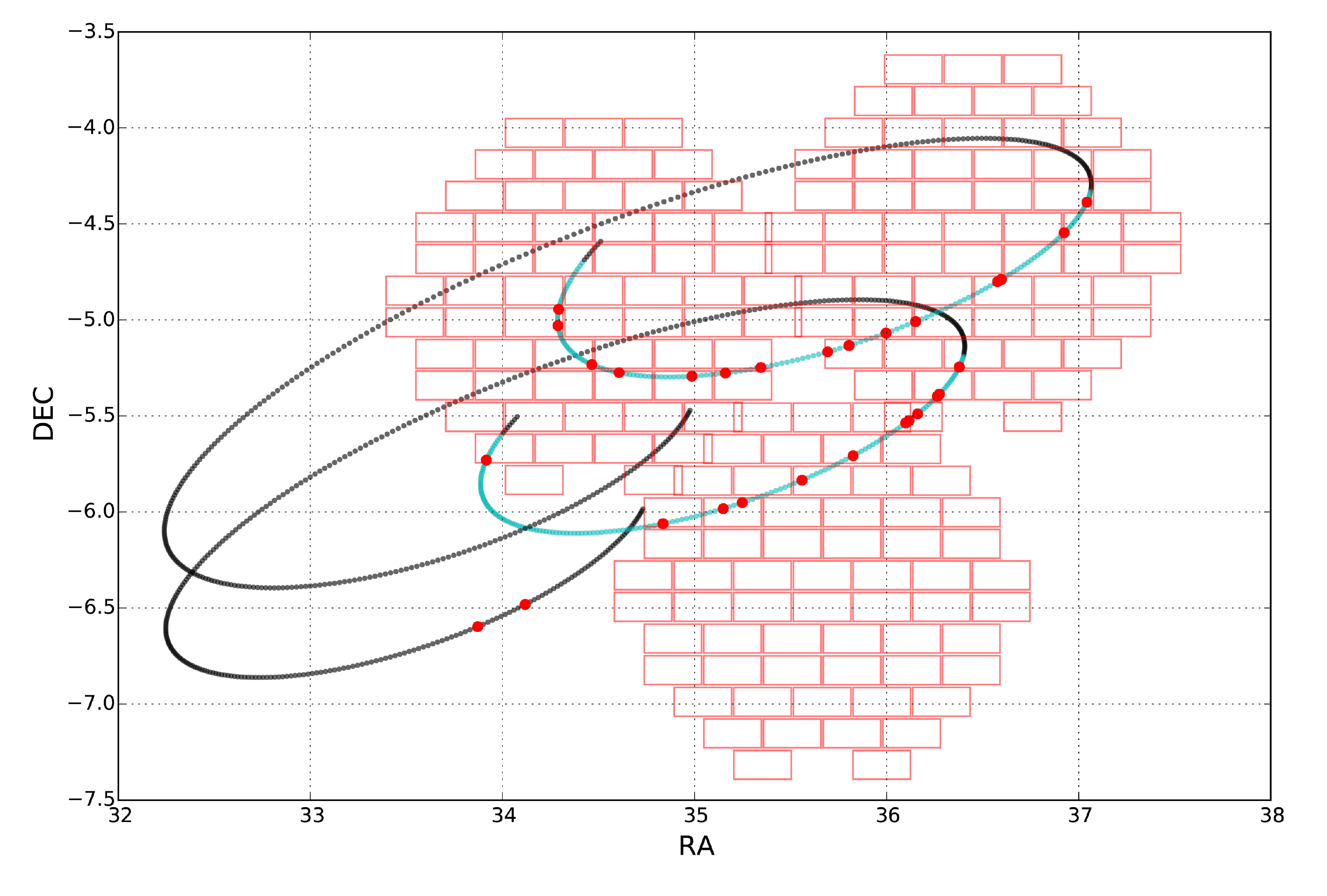}
\caption{\label{fig:trajectory} Observed trajectories of \QO\ (lower curve) and \QP\ (upper curve) relative to the three DES XMM-LSS supernova fields over the first two DES observing seasons. The cyan portions of the trajectories encompass the 2014-15 DES observing season. Large dots indicate nights on which the objects were detected. Each rectangle represents one DECam CCD. \QO\ was also
detected in wide-survey exposures outside of the supernova fields on two nights in October, 2013.}
\end{center}
\end{figure}

\QO\ and \QP\ were initially detected at apparent $r$-band magnitudes of 23.3 and 23.9 in the deep XMM-LSS field on Aug.\ 21, 2014, at 
ecliptic latitudes of $-18.5^{\circ}$ and $-18.0^{\circ}$ respectively.  \QO\ was observed 69 times on 12 different nights
through Jan.\ 29, 2015, while \QP was observed 62 times on 13 nights through Jan.\ 22, 2015. In addition, \QO\ was recovered in five 90-second wide-survey exposures from
Oct.\ 2 and 13, 2013. The trajectories of these two objects are shown in Figure~\ref{fig:trajectory}. Orbital elements and other properties of these two
objects are summarized in Table~\ref{tab:parameters}. The 2013 observations of \QO\ increase its arc length by nearly a year and substantially reduce the uncertainty in the orbit.
A summary of these objects' properties in comparison with the other known Neptune Trojans is shown in Table~\ref{tab:alltrojans}.

\begin{deluxetable}{lll}
\tabletypesize{\scriptsize}
\tablecaption{Orbital elements and other properties of the two new Trojans. \label{tab:parameters} }
\tablewidth{0pt}
\tablehead{
\colhead{Parameter} & \colhead{\QO}  & \colhead{\QP} 
}
\startdata
$a$~(AU) & $30.102 \pm 0.001$ & $30.108\pm 0.011$  \\
$e$ & $0.1046 \pm 0.0004$ & $0.067 \pm 0.002$ \\
$i$ (deg) & $18.8316 \pm0.0003$ & $19.405\pm  0.001$  \\
$\omega$ (deg) & $113.1 \pm 1.0$ & $1.74 \pm 0.67$  \\
$\Omega$ (deg) & $106.998 \pm  0.004$ & $96.538 \pm 0.007$ \\
Perihelion date & 1937/11/28 $\pm$ 176d & 2043/4/16 $\pm$ 79d  \\
Epoch JD & 2457000.5 & 2457000.5 \\
\pbox{3cm}{Libration\\  \hspace*{4pt} period (yr)} & $9074\pm 3$ & $9114\pm 5$ \\
\pbox{3cm}{Libration\\ \hspace*{4pt} amplitude (deg.)} & $10.9\pm0.1$ & $7.8\pm1.3$ \\
Arc length (days) & 484 & 154 \\
Apparent mag (r) & 23.3 & 23.9 \\
$H_{V}$ & 8.2 & 9.1 \\
$r-i$ (mag.) & $0.21\pm 0.10$& $0.20\pm 0.06$ \\
$i-z$ (mag.) & $0.01\pm 0.11$& $0.08\pm 0.08$ \\  
Diameter (km)\tablenotemark{a} & 120 & 70 \\
\enddata
\tablenotetext{a}{Assuming 5\% albedo.}
\end{deluxetable}

\begin{deluxetable}{llllllcll} 
\tabletypesize{\scriptsize}
\tablecaption{Properties of Neptune Trojans \label{tab:alltrojans} }
\tablewidth{0pt}
\tablehead{
\colhead{Name} & \colhead{$a$ (AU)\tablenotemark{a}} & \colhead{$e$} & \colhead{$i$ (deg.)} &  \colhead{$H_{V}$} & \colhead{L} & \colhead{diam. (km)\tablenotemark{b}} & 
\colhead{Libration ampl. (deg.)\tablenotemark{c}}  & \colhead{Libration per. (yr.)} 
}
\startdata
2001 QR$_{322}$ & 30.207 & 0.027 &1.3 & 8.2 & L4 & 140 & $25.5^{+0.4}_{-0.8}$ & $9200\pm 3$ \\
2004 UP$_{10}$ & 30.130 & 0.024 & 1.4 & 8.8 & L4 & 100 & $10.8^{+1.0}_{-0.3}$ & $8874 \pm 8$\\
2005 TN$_{53}$ & 30.127 & 0.067 & 25.0 & 9.0 & L4 & 80 & $8.7^{+0.3}_{-0.5}$ &$9428\pm 6$ \\
2005 TO$_{74}$ & 30.120 & 0.055 & 5.3 &8.5 & L4 & 100 & $9.2^{+0.2}_{-0.5}$ & $8822\pm10$\\
2006 RJ$_{103}$ & 30.045 & 0.032 & 8.2 &7.5 & L4 & 180 & $6.3^{+0.1}_{-0.3}$ & $8858\pm 10$ \\
2007 VL$_{305}$ & 30.079 & 0.065 & 28.1 & 8.0 & L4 & 160 & $14.2^{+0.03}_{-0.10}$ & $9619\pm 8$\\
2008 LC$_{18}$ & 29.922 & 0.085 & 27.6 & 8.4 & L5 & 100 & $16.4^{+1.3}_{-1.1}$ & $9516\pm 7$\\
2011 HM$_{102}$ & 30.059 & 0.080 & 29.4 & 8.1 & L5 & 140 & $9.8^{+0.4}_{-0.4}$ & $9545 \pm 4$\tablenotemark{d}\\
2012 UV$_{177}$ & 30.032 & 0.074 & 20.8 & 9.2 & L4 & 80 &$13$\tablenotemark{e} & $9334\pm 10$\\
2014 QO$_{441}$ & 30.089 & 0.105 & 18.8 & 8.2 & L4 & 120 & $10.9^{+0.1}_{-0.1}$  & $9074 \pm 3$\\
2014 QP$_{441}$ & 30.108 & 0.067 & 19.4 &9.1 & L4 & 80 & $7.9^{+1.3}_{-1.3}$ &$9114\pm 5$ \\
\enddata
\tablenotetext{a}{Barycentric osculating elements at epoch 2014/12/09.}
\tablenotetext{b}{Assuming 5\% albedo.}
\tablenotetext{c}{Values for first 8 objects from \citet{Parker2015}.}
\tablenotetext{d}{\citet{Parker2013}}
\tablenotetext{e}{\citet{Alexandersen2014}}
\end{deluxetable}

\section{Dynamical Stability}
\label{sec:stability}

\begin{figure}[htbp]
\begin{center}
\includegraphics[width=\columnwidth]{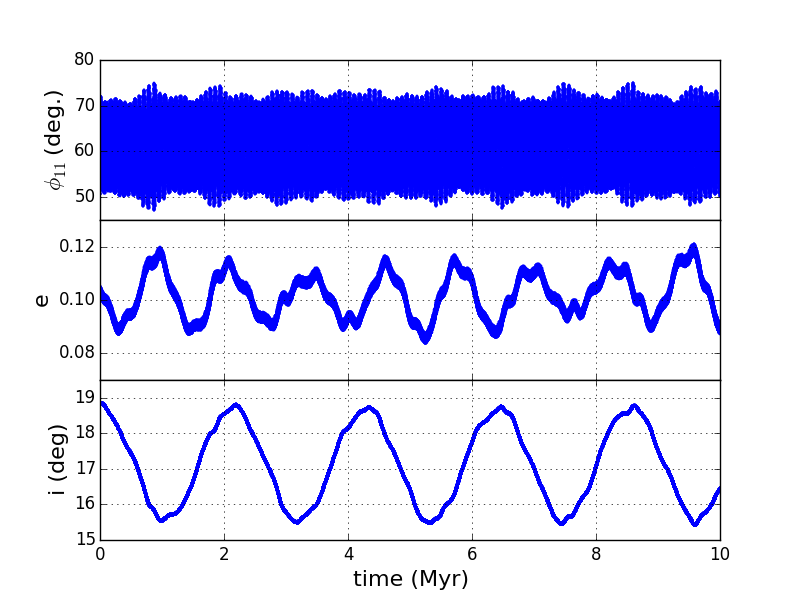}
\includegraphics[width=\columnwidth]{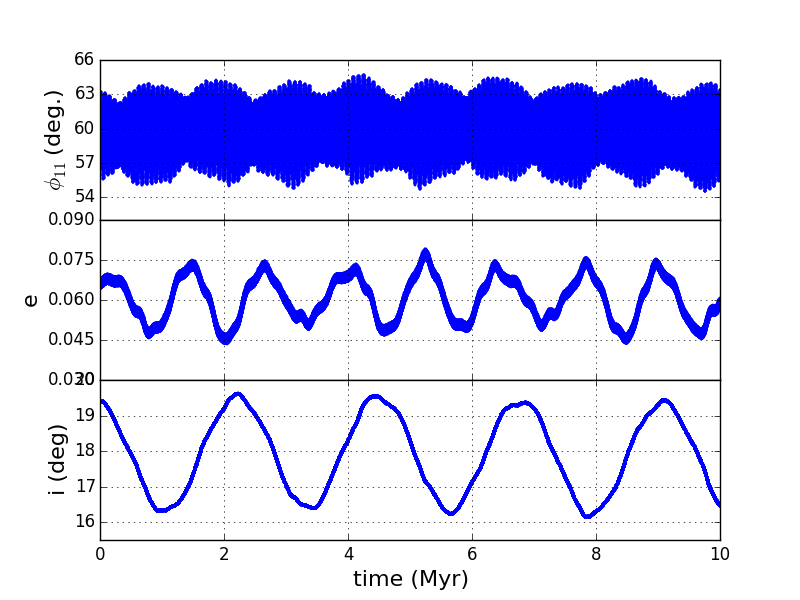}
\caption{\label{fig:10Myr-aei} Evolution of the resonant angle $\phi_{11}$ and the orbital parameters $e$ and $i$ over a 10Myr integration for the best-fit orbits of \QO\ (top) and \QP\ (bottom). This pattern
continues stably for each of the 1000 clones analyzed.}
\end{center}
\end{figure}

To investigate the long-term behavior of the orbits, we have simulated swarms of 1000 clones each of \QO\ and \QP. The initial orbital elements of each clone were drawn from a multivariate normal distribution using the full six-dimensional covariance matrix obtained from the fit to the observations. We followed
each clone for 10~Myr in the presence of the four giant planets using the hybrid symplectic algorithm in the \textsc{Mercury6} $N$-body integrator \citep{Chambers1999}. Each of the clones remained in a bound orbit about L4, evolving stably in $aei$-space. An example is shown in Figure~\ref{fig:10Myr-aei}.

\begin{figure}[htbp]
\begin{center}
\includegraphics[width=\columnwidth]{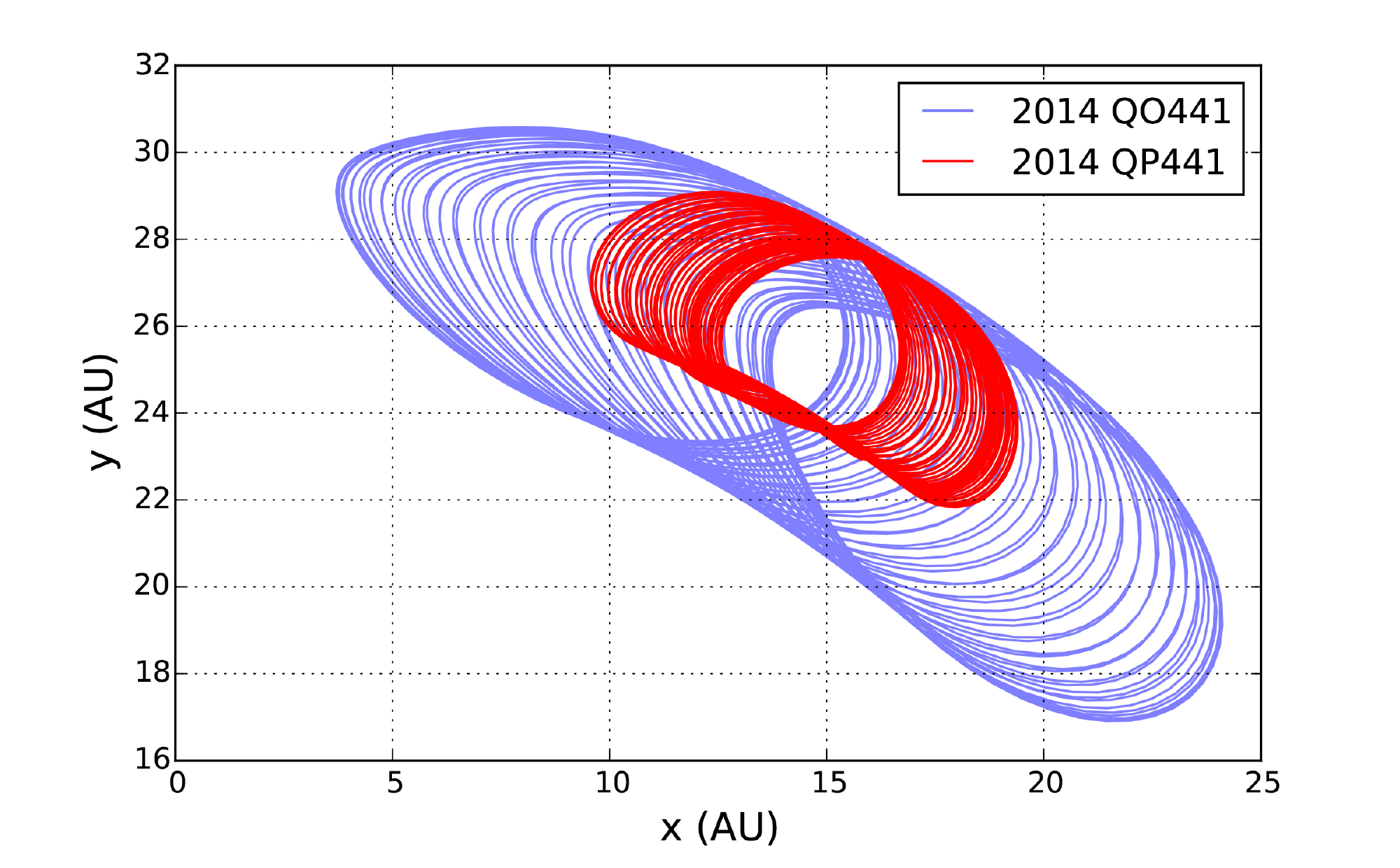}
\caption{\label{fig:corotating} Orbital motion of \QO\ (blue) and \QP\ (red) in the ecliptic plane over the course of one full $\sim 9100$-year libration period 
as seen from a frame rotating counterclockwise at the speed of Neptune. In this coordinate system the Sun is at (0,0) and Neptune lies along the $x$-axis
between $x=28.9107$ and $x=30.3271$. Each individual sub-cycle of the full pattern corresponds to one 165-year orbit. }
\end{center}
\end{figure}

To compute the libration amplitudes and periods we follow the procedure of \citet{Parker2013} and \citet{Parker2015}. Using the same 1000 clones of each object, we define the half-peak rms libration amplitude for each clone to be
\begin{equation}
	L_{11} = \left(\frac{2}{N}\sum_{i=1}^{N}(\phi_{i}-\langle\phi\rangle)^{2}   \right)^{1/2},
\label{eqn:lib-amplitude}
\end{equation}
where $i$ runs over the $N$ samples of each clone during the 10~Myr integration. The resonant argument $\phi_{11} = \lambda_{N}-\lambda_{T}$, where $\lambda_{N}$ and
$\lambda_{T}$ are the mean longitude of Neptune and the Trojan respectively \citep{Chiang2003}. The mean longitude computed at each epoch of the integration is
\begin{equation}
\lambda = M + \Omega + \omega,
\label{eqn:mean-longitude}
\end{equation}
where $M$ is the mean anomaly, $\Omega$ is the longitude of the ascending node, and $\omega$ is the argument of perihelion. 
 Libration periods are obtained from a spectral decomposition of $\phi(t)$, and errors
are obtained from the rms scatter among the clones. We find that \QO\ librates with an rms amplitude of $10.9\pm 0.1^{\circ}$ and a period of $9074\pm 3$ years, while for
\QP\ we find a libration amplitude of $7.8\pm1.3^{\circ}$ and a period of $9114\pm 5$ years. The orbital motion of both Trojans in the co-rotating Neptunian frame over the course of one full libration period is shown in Figure~\ref{fig:corotating}.

In addition to the $\sim$9100-year libration periods, we
observe oscillations in eccentricity and inclination on 1-2 Myr timescales, which are a by-product of the long-term exchange of angular momentum between Jupiter and Saturn. To investigate these further we have continued 32 of the integrations for each object for 1~Gyr. We observe that these oscillations continue, but do not grow in amplitude, over the full timespan of the integrations for \QP. Five of the 32 clones of \QO\ did not survive the full 1~Gyr integration, with the earliest instability occurring between 500-600 Myr. Thus \QP\ appears to be completely stable in its present configuration, while \QO\ possesses a dynamical half-life on the order of 4~Gyr. Both objects could therefore be primordial members of the disk into which Neptune migrated.

\section{Physical Characterization}
\label{sec:physical}
The measured $r-i$ and $i-z$ colors shown in Table~\ref{tab:parameters} indicate that both objects are slightly red. The derived $V-R$ magnitudes of $\approx 0.44\pm 0.11$ are consistent with 
previously reported colors of the L4 and L5 Neptune Trojans \citep{Parker2013} as well as Jupiter Trojans and neutral Centaurs, and are slightly bluer than classical KBOs. This similarity may 
indicate a common formation mechanism and history, although future near infrared photometry may reveal otherwise unapparent population differences as is known to be the case with Jupiter Trojans \citep{Wong2014}.

We examined the light curves of each object to search for periodic behavior indicative of rotation or a possible binary system. We carried out this analysis using both the Phase Dispersion Minimization \citep{Stellingwerf1978} and generalized Lomb-Scargle periodogram \citep{Scargle1982, Zechmeister2009} techniques. We found no evidence for modulations on timescales between 0.5-48 hours, leading us to conclude that within our photometric sensitivity both \QO\ and \QP\ are relatively round and featureless.

\section{Summary}
\label{sec:summary}

This paper reports the discovery of \QO\ and \QP\, which represent the eighth and ninth members of the L4 family of Neptune Trojans. They appear to be dynamically stable on $>\rm Gyr$ timescales and have 
colors similar to other Trojans. Their high inclinations lend further support to the idea that Neptune Trojans are a dynamically ``hot'' population that may have been excited
prior to capture by Neptune \citep{Parker2015}. The Dark Energy Survey will have further opportunities to observe \QO\ during the 2015-16 and 2016-17 seasons, when it will again enter the deep XMM-LSS supernova field. \QP\ will also be visible in this 
field during the 2015-16 season, then will briefly pass through one of the Stripe-82 fields during the 2016-17 and 2017-18 seasons. Further measurements of this object will improve the estimate of its
dynamical lifetime.

The existence of the Neptune Trojans, along with other small bodies in
the Solar System, poses a set of interesting dynamical problems.
These bodies could have been placed in their current orbits at the epoch of
solar system formation, or perhaps more recently.  In the former
scenario, the orbits must be stable over the $\sim4.6$ Gyr age of the
Sun. Sufficiently large perturbations to the orbit of Neptune, or the
other giant planets, could lead to the removal of Trojans from their
librations about the Lagrange points (e.g., Figure~\ref{fig:10Myr-aei} shows that small
perturbations of Neptune due to the other planets lead to changes in
the resonant angle). As a result, the long-term stability of these
orbits implies upper limits on the degree of disruption suffered by
the solar system.  On the other hand, if the placement of the Trojans
into their currently observed orbits is more recent, then the
mechanism that populates such orbits must be understood. These
issues thus present a rich set of open dynamical questions.

Although this work has emphasized the detailed study of these two new Trojans, an additional 20 Kuiper Belt and other trans-Neptunian
objects discovered in our search of the DES supernova fields have been given provisional designations by the Minor Planet Center. The most notable of these is 2013~RF$_{98}$, which has a semi-major axis of 325 AU and a perihelion of 36 AU, making it one of the longest-period trans-Neptunians known. These objects will be described in forthcoming work. 

With three years remaining in the Dark Energy Survey, we can confidently anticipate that several dozen new trans-Neptunian objects, and perhaps including more Neptune Trojans, will enter the supernova fields and be discovered by the techniques described in this paper. But the supernova fields represent less than 1\% of the full 5000 sq.\ deg.\ wide survey area. Although the sparse cadence with which a given wide survey region is observed---typically 2-4 times per season in each of five filters---makes the identification of slow-moving transients more challenging, a search in the wide survey fields is likely to result in hundreds of TNOs to a limiting magnitude of $r\sim 23.8$. Moreover, the wide survey footprint includes a small region near the ecliptic, but is distributed primarily over higher ecliptic latitudes, an area not well covered by previous TNO searches. This makes DES particularly well suited for the detection of the higher-inclination, ``hot'' TNO populations. DES's combination of area and depth suggests that DES can discover 6-10 times more such objects than any previous survey. Because the detached / inner Oort cloud population is even less likely to be confined to the ecliptic region, we expect DES to have at least an order of magnitude more sensitivity to this very interesting and rare group of objects compared to prior searches, shedding further light on the processes that shaped the solar system. 

\section{Acknowledgements}
We are grateful for the extraordinary contributions of our CTIO colleagues and the DES Camera, Commissioning and
Science Verification teams for achieving excellent instrument and telescope conditions that have made this work possible.
The success of this project also relies critically on the expertise and dedication of the DES Data Management organization.
Funding for the DES Projects has been provided by the U.S. Department of Energy, the U.S. National Science Foundation, the Ministry of Science and Education of Spain, the Science and Technology Facilities Council of the United Kingdom, the Higher Education Funding Council for England, the National Center for Supercomputing Applications at the University of Illinois at Urbana-Champaign, the Kavli Institute of Cosmological Physics at the University of Chicago, Financiadora de Estudos e Projetos, Funda{\c c}{\~a}o Carlos Chagas Filho de Amparo {\`a} Pesquisa do Estado do Rio de Janeiro, Conselho Nacional de Desenvolvimento Cient{\'i}fico e Tecnol{\'o}gico and the Minist{\'e}rio da Ci{\^e}ncia e Tecnologia, the Deutsche Forschungsgemeinschaft and the Collaborating Institutions in the Dark Energy Survey.

The Collaborating Institutions are Argonne National Laboratory, the University of California at Santa Cruz, the University of Cambridge, Centro de Investigaciones Energeticas, Medioambientales y Tecnologicas-Madrid, the University of Chicago, University College London, the DES-Brazil Consortium, the Eidgen{\"o}ssische Technische Hochschule (ETH) Z{\"u}rich, Fermi National Accelerator Laboratory, the University of Edinburgh, the University of Illinois at Urbana-Champaign, the Institut de Ciencies de l'Espai (IEEC/CSIC), the Institut de Fisica d'Altes Energies, Lawrence Berkeley National Laboratory, the Ludwig-Maximilians Universit{\"a}t and the associated Excellence Cluster Universe, 
the University of Michigan, the National Optical Astronomy Observatory, the University of Nottingham, The Ohio State University, the University of Pennsylvania, the University of Portsmouth, SLAC National Accelerator Laboratory, Stanford University, the University of Sussex, and Texas A\&M University.

This paper has gone through internal review by the DES collaboration.

 \bibliography{apj-jour,trojans}

\label{lastpage}
\end{document}